\documentclass[aps,prd,showpacs]{revtex4}
\usepackage[dvips]{graphicx}

\begin{document}

\title{Instability of a gapless color superconductor with respect to
inhomogeneous fluctuations}

\author{Kei Iida and Kenji Fukushima}
\affiliation{RIKEN BNL Research Center, Brookhaven National Laboratory,
Upton, New York 11973, USA}
\date{\today}

\begin{abstract}
     We systematically apply density functional theory to determine 
the kind of inhomogeneities that spontaneously develop in a homogeneous 
gapless phase of neutral two-flavor superfluid quark matter.  We consider
inhomogeneities in the quark and electron densities and in the phases and 
amplitude of the order parameter.  These inhomogeneities are expected to
lead the gapless phase to a BCS-normal coexisting phase, a 
Larkin-Ovchinnikov-Fulde-Ferrell (LOFF) state with phase oscillations alone, 
and a LOFF state with amplitude oscillations.  We find that which of them the
homogeneous system tends towards depends sensitively on 
the chemical potential separation between up and down quarks
and the gradient energies.

\end{abstract}


\maketitle

\section{Introduction}
\label{sec:intro}

     It has been noted since the seminal work in Refs.\ \cite{barrois,BL} 
that the quark-quark interaction in the color antitriplet channel is attractive
and drives a Cooper pairing instability in quark matter in the limit of high 
density.  In this limit, the ground state for three-flavor quark matter is a 
homogeneous color-flavor locked state in which all nine quarks, associated with
three colors and three flavors, are gapped \cite{ARW}.  However, color 
superconducting states, if occurring in compact stars, would not be necessarily
homogeneous.  This is because a separation of the Fermi surface develops 
between paired quarks in a realistic situation characterized by nonzero strange
quark mass, color and electric charge neutrality, and weak equilibrium.  Once 
the separation becomes comparable to 
the gap magnitude, a usual BCS state can be energetically unfavorable because 
of the inevitable increase in the loss of the quark kinetic energy. 
Even in this situation, it is possible to consider other homogeneous 
paired states such as a gapless state in which quark quasiparticles 
which are gapped in the absence of the Fermi surface separation become gapless 
\cite{GLW,SH,AKR}, a color-flavor locked state with condensation of collective
modes carrying the same quantum number as mesons \cite{BSKR}, and a paired 
state with nonspherical Fermi surfaces \cite{MS}.  Importantly, these 
homogeneous states are not always stable against inhomogeneities.  The most 
remarkable example is a chromomagnetic instability of the gapless states which 
is characterized by negative Meissner masses squared \cite{HS2,CN}; 
this signifies that fluctuations in the gluon fields or, equivalently,
the phases of the order parameter \cite{GR}, develops spontaneously.  [Note 
that in an overall neutral system of charged fermions, the gapless states can 
be stable against homogeneous change in the gap magnitude in contrast to the 
case of neutral Fermi systems (see Ref.\ \cite{Forbes} for an exception).]  
So far, many states involving inhomogeneities in the quark densities and/or 
the order parameter \cite{Shov} have been proposed for an eventual stable 
state after the onset of the chromomagnetic instability, but a systematic 
energy comparison between them and even a systematic instability analysis with
respect to various possible inhomogeneities remain to be performed.

     In this paper, as a first step towards such a systematic analysis,
we utilize density functional theory to investigate instabilities of a 
two-flavor homogeneous gapless state with respect to inhomogeneous 
fluctuations in the quark and electron densities and in the phases and 
amplitude of the order parameter.
Here we do not address what the ground state is, but all we
can clarify is to identify inhomogeneities that grow spontaneously
in the homogeneous system, from which we can anticipate
what kind of inhomogeneous state is likely to be realized.
Possibly, the growth of fluctuations in the 
phases and/or amplitude of the order parameter would end up with a 
Larkin-Ovchinnikov-Fulde-Ferrell (LOFF) state \cite{LOFF} in which periodic 
spatial oscillations occur in the phases and/or amplitude, while the 
simultaneous growth of fluctuations in
the density difference between up and down quarks and 
the gap magnitude could lead to a BCS-normal coexisting phase \cite{BCR,RR}.  
We take into account finite size corrections such as the energies arising from 
the density and gap gradients and the electrostatic and color Coulomb
energies to deduce the structure of the coexisting phase.  We find that 
electric charge screening, which is automatically included in the present
density functional analysis, is crucial to such deduction.  
The coexisting phase and the LOFF phase with amplitude modulations
are related with each other in the sense that both phases involve
nonzero amplitude and density modulations.  As we will see later, it is 
the competition between the density and gap gradient energies 
that determines which of these the system prefers
to go to when the system is unstable with respect to fluctuations in
the gap magnitude.

     In Sec.\ \ref{sec:homo}, we summarize the bulk properties of 
homogeneous color superconducting states in two-flavor neutral quark matter
at zero temperature.  Section \ref{sec:inst} is devoted to constructing 
density functional theory, by including not only the bulk properties summarized
in Sec.\ \ref{sec:homo} but also inhomogeneities in the order parameter 
and the densities of the constituents, and to classifying possible 
instabilities of the gapless state.   In Sec.\ \ref{sec:pot}, we 
calculate the effective potential for the gap magnitude and thereby
clarify how the type of instabilities changes with the Fermi surface 
separation and a parameter characterizing the density gradient
energy.  Our conclusions are given in Sec.\ \ref{sec:conc}.  
We use units in which $\hbar=c=k_{B}=1$.

\section{Homogeneous phases}
\label{sec:homo}

     In this section we summarize the bulk properties of two-flavor neutral 
$\beta$ equilibrated quark matter at zero temperature by using a simple BCS 
type model.  This is instructive for a later stability analysis, which 
requires the second derivatives of the thermodynamic potential density with 
respect to the phases and magnitude of the order parameter and the quark and 
electron densities.

     We consider uniform $ud$-flavor quark matter of baryon chemical 
potential $3\mu$ and zero temperature.  We assume that the system is
in weak equilibrium with the gas of electrons of chemical potential $\mu_e$ 
and has zero net color and electric charge.  We neglect quark and electron 
masses, and chiral and meson condensates.  For a normal state, we simply
adopt the ideal gas form of the thermodynamic potential density,
\begin{equation}
\Omega_{\rm normal}=-\frac{1}{12\pi^2}\sum_{i=e,q}\mu_i^4,
\label{enormal}
\end{equation}
where $q\equiv fa$ stands for the set of color $a=R,G,B$ and flavor $f=u,d$, 
and $\mu_q$ is the chemical potential of $q$ quarks.  In this case, $\beta$ 
equilibrium and charge neutrality lead to
\begin{equation}
   \mu_q= \mu - Q_q \mu_e,
\label{muq}
\end{equation}
where $Q_q$ is the electric charge of $q$ quarks.  The number density of $i$ 
particles are then given by the corresponding chemical potentials $\mu_i$ as
\begin{equation}
   n_i=\frac{\mu_i^3}{3\pi^2}.
\label{ninm}
\end{equation}

     Let us then consider a two-flavor color superconducting (2SC) state
in which Cooper pairing occurs in a $J^P=0^+$, isoscalar, and $RG$-color 
antitriplet channel.  For a pairing interaction, we adopt a contact
interaction,
\begin{equation} 
   -G_D(\bar{\psi}_{f a}
   P_{a b}^{f h}\psi^C_{h b}) (\bar{\psi}^C_{f' a'}
   P_{a'b'}^{f'h'}\psi_{h'b'}),
\end{equation}
where $P_{ab}^{fh}=i\gamma_5\epsilon_{fh}\epsilon_{abB}$, $\psi_{fa}$ is
the quark spinor of color $a$ and flavor $f$, and 
$\psi_{fa}^C=i\gamma^2\gamma^0 {\bar\psi}_{fa}^T$.
Here we assume that this pairing interaction is cut off unless the transfer of
the relative momentum of paired quarks is smaller than $\Lambda$ in magnitude.
In some sense this cutoff can be considered to simulate the instanton form 
factor \cite{RSSV}, which allows us to solve the gap equation in the 
same way as the original BCS case in which the cutoff is given by the 
Debye frequency.  We take $\Lambda$ as 300~MeV, which is small compared with 
a typical value of the Fermi energy of order 500~MeV.  Hereafter we will 
often assume $\Delta \ll \Lambda$; this is reasonable for a typical range of 
$\Delta$ of 0--100~MeV.

     Within the mean-field approximation, the pairing gap $\Delta$ is related 
to the diquark condensate as
\begin{equation}
 \Delta =2G_D\langle\bar{\psi}_{fa}P_{ab}^{fh}\psi^C_{hb}\rangle,
\label{eq:delta}
\end{equation}
with the ensemble average $\langle\cdots\rangle$.  Taking note of this 
relation, it is straightforward to write down the thermodynamic potential 
density as \cite{SH}
\begin{eqnarray}
\Omega_{\rm 2SC}\!&=&\!-\frac{1}{12\pi^2}\!\sum_{i=e,uB,dB}\!\mu_i^4
  -\frac{1}{12\pi^2}\!\!\sum_{i=uR,dR,uG,dG}\!\!{\bar\mu}^4
  +\frac{\Delta^2}{4G_D}
  -\frac{1}{2\pi^2}\!\!\sum_{i=uR,dR,uG,dG}\!\!
   \int_{{\bar\mu}-\Lambda}^{{\bar\mu}+\Lambda}dp p^2
   \left[\sqrt{(p-{\bar\mu})^2+\Delta^2}-|p-{\bar\mu}|\right]
 \nonumber \\ 
   &\simeq& -\frac{1}{12\pi^2}\sum_{i=e,uB,dB}\mu_i^4
  -\frac{1}{12\pi^2}\sum_{i=uR,dR,uG,dG}{\bar\mu}^4  +\frac{\Delta^2}{4G_D}
  -\frac{{\bar\mu}^2\Delta^2}{2\pi^2}\sum_{i=uR,dR,uG,dG}
   \left(\frac12+\ln\frac{2\Lambda}{\Delta}\right),
  \label{omega2SC}
\end{eqnarray}
where 
\begin{equation}
  {\bar\mu}\equiv\frac{\mu_{uR}+\mu_{dG}}{2}=\frac{\mu_{dR}+\mu_{uG}}{2},
\label{eq:barmu}
\end{equation}
and $p$ is the quasiparticle momentum associated with paired quarks.
We have assumed $\Delta\ll\Lambda$ in the approximate estimate.
Weak equilibrium and color and electric charge neutrality ensure 
\begin{equation}
 \mu_{fa}=\mu-Q_f \mu_e+Q^{\alpha=8}_a \mu_8,
 \label{mufa}
\end{equation}
where $Q^{\alpha=8}_a\equiv(1/3,1/3,-2/3)$ is the color charge for gluon color 
index $\alpha=8$, and $\mu_8$ is the associated color chemical potential.
The chemical potentials $\mu_e$ and $\mu_8$ are determined by the 
neutrality conditions,
\begin{equation}
   \frac{\partial\Omega_{\rm 2SC}}{\partial\mu_e}
  =\frac{\partial\Omega_{\rm 2SC}}{\partial\mu_8}=0.
  \label{neut}
\end{equation}
From these conditions, one can show that in the 2SC state, $\mu_e$ is related 
to $\mu$ and $\mu_8$ as $\mu_e=(3\mu-2\mu_8)/5$, while $\mu_8\approx0.0854\mu$ 
up to leading order in $\Delta/\mu$ \cite{SH}.

     We then calculate the gap and the number densities from 
$\Omega_{\rm 2SC}[\Delta; \mu_i]$ given by Eq.\ (\ref{omega2SC}).  Variation 
of $\Omega_{\rm 2SC}$ with respect to the gap gives rise to the gap equation,
\begin{equation}
  0=\frac{\partial\Omega_{\rm 2SC}}{\partial \Delta}
   =\frac{\Delta}{2G_D}-\frac{4{\bar\mu}^2 \Delta}{\pi^2}
    \ln\frac{2\Lambda}{\Delta}.
  \label{gapeq2SC}
\end{equation}
This is identical with the form of the usual BCS gap equation.  We define the 
gap satisfying Eq.\ (\ref{gapeq2SC}) as $\Delta=\Delta_{\rm 2SC}$.  The number
densities for $i=e,uB,dB$ are still given by Eq.\ (\ref{ninm}), while for
$i=uR,uG,dR,dG$ we obtain, up to leading order in $\Delta/\Lambda$,
\begin{equation}
   n_i\simeq\frac{{\bar\mu}^3}{3\pi^2}+\frac{{\bar\mu}\Delta^2}{\pi^2}
      \ln\frac{2\Lambda}{\Delta}.
    \label{ni2SC}
\end{equation}
Obviously, the neutrality constraints (\ref{neut}) can be rewritten
in terms of $n_i$ as
\begin{equation}
   n_e=\sum_{fa}Q_f n_{fa},\qquad 0=\sum_{fa}Q^{\alpha=8}_a n_{fa}.
   \label{neut2}
\end{equation}

    The 2SC state is relevant only when the separation of the Fermi surface
between the $u$ and $d$ quarks,
\begin{equation}
  \delta\mu\equiv\frac{\mu_{dG}-\mu_{uR}}{2}
                =\frac{\mu_{dR}-\mu_{uG}}{2}=\frac{\mu_e}{2},
\end{equation}
is smaller than the gap $\Delta_{\rm 2SC}$ \cite{SH}.  This is because the 
smallest quasiparticle gap associated with paired quarks vanishes once the 
separation reaches $\Delta_{\rm 2SC}$.  In fact, according to the usual
Bogoliubov method, the corresponding quasiparticle energy $E_{-}(p)$ reads 
\begin{equation}
  E_{-}(p)=\sqrt{(p-{\bar\mu})^2+\Delta^2}-\delta\mu,
  \label{equasi}
\end{equation}
and for $\delta\mu=\Delta=\Delta_{\rm 2SC}$, $E_{-}({\bar\mu})=0$.  Note that 
in the 2SC state the gap magnitude and thus the condensation energy is
independent of $\delta\mu$ except for the slight dependence on 
$\delta\mu$ through ${\bar\mu}$ in Eq.~(\ref{gapeq2SC}).  This is because 
quarks forming Cooper pairs are located in the middle of the Fermi surfaces 
as in Eq.~(\ref{eq:barmu}), rather than around the respective Fermi surfaces.

    For $\delta\mu < \Delta_{\rm 2SC}$, one can find another branch of the 
homogeneous solution to the gap equation in which the gap magnitude is
not only smaller than $\Delta_{\rm 2SC}$ but also the 
separation $\delta\mu$.  This state, hereafter referred to as the gapless 2SC 
state, is characterized by the presence of gapless quasiparticle modes 
associated with paired quarks.  This is similar to a state originally 
predicted by Sarma \cite{Sarma} for superconductors in external magnetic 
fields.  The corresponding thermodynamic potential can be described, simply 
by adding to Eq.\ (\ref{omega2SC}) an energy contribution from gapless 
quasiparticle modes of energy $E_{-}(p)$, as
\begin{eqnarray}
\Omega_{\rm g2SC}&=&-\frac{1}{12\pi^2}\!\sum_{i=e,uB,dB}\!\mu_i^4
  -\frac{1}{12\pi^2}\!\!\sum_{i=uR,dR,uG,dG}\!\!{\bar\mu}^4  
  +\frac{\Delta^2}{4G_D}
  -\frac{1}{2\pi^2}\!\!\sum_{i=uR,dR,uG,dG}\!\!\int_{{\bar\mu}
  -\Lambda}^{{\bar\mu}+\Lambda}dp p^2 
  \left[\sqrt{(p-{\bar\mu})^2+\Delta^2}-|p-{\bar\mu}|\right]
  \nonumber \\ && 
 +\frac{1}{2\pi^2}\sum_{i=uR,dR,uG,dG} \int_{\mu^-}^{\mu^+}dp p^2 E_{-}(p),
  \label{omegag2SC}
\end{eqnarray}
where $\mu^{\pm}\equiv{\bar\mu}\pm\sqrt{(\delta\mu)^2-\Delta^2}$.  Then,
the gap equation reads
\begin{eqnarray}
  0=\frac{\partial\Omega_{\rm g2SC}}{\partial \Delta}
   &=&\frac{\Delta}{2G_D}-\frac{4{\bar\mu}^2 \Delta}{\pi^2}
    \ln\frac{2\Lambda}{\Delta}
   \nonumber \\ &&+2\frac{\Delta}{\pi^2}
      \left[(2{\bar\mu}^2-\Delta^2)
      \ln\frac{\sqrt{(\delta\mu)^2-\Delta^2}+\delta\mu}{\Delta}
      +\delta\mu\sqrt{(\delta\mu)^2-\Delta^2}
      \right]. 
      \label{gapeqg2SC}
\end{eqnarray}
By retaining a term of leading order in $\Delta/{\bar\mu}$ in the contribution
from the gapless modes, we obtain an approximate solution as
\begin{equation}
 \Delta\simeq\sqrt{\Delta_{\rm 2SC}(2\delta\mu-\Delta_{\rm 2SC})},
    \label{gapg2SC}
\end{equation}
which has the same form as that derived by Sarma \cite{Sarma}.  Note that the 
gap magnitude is explicitly dependent on $\delta\mu$ in contrast to the case 
of the 2SC state.

    The number densities of quarks and electrons can be determined
by the derivatives of $\Omega_{\rm g2SC}$ with respect to
the chemical potentials.  The number densities
for $i=e,uB,dB$ are again given by Eq.\ (\ref{ninm}), while for
$i=uR,uG,dR,dG$ we obtain, up to leading order in $\Delta/\Lambda$,
\begin{eqnarray}
   n_i &\simeq& \frac{{\bar\mu}^3}{3\pi^2}+\frac{{\bar\mu}\Delta^2}{\pi^2}
      \ln\frac{2\Lambda}{\Delta}
     \nonumber \\
     &\pm&\frac{1}{3\pi^2}
      \sqrt{(\delta\mu)^2-\Delta^2}[3{\bar\mu}^2+(\delta\mu)^2-\Delta^2]
      -\frac{{\bar\mu}}{\pi^2}
    \left[\Delta^2\ln\frac{\sqrt{(\delta\mu)^2-\Delta^2}+\delta\mu}{\Delta}
        -\delta\mu \sqrt{(\delta\mu)^2-\Delta^2} \right],
    \label{nig2SC}
\end{eqnarray}
where the upper and lower sign in the right side are taken for $d$ quarks and 
$u$ quarks, respectively.   

     It is instructive to note that the gapless state is stable against small 
homogeneous change in the gap magnitude in charged Fermi systems
because of the neutrality constraints \cite{SH}, although it is generally 
unstable in neutral Fermi systems. (In fact, without charge, at given 
$\delta\mu$ below $\Delta_{\rm 2SC}$, the gap magnitude and thus the 
condensation energy would be larger for the 2SC state than for the gapless
state.)  By substituting the above-obtained densities into Eq.\ (\ref{neut2}), 
one obtains an approximate relation between $\Delta$ and $\delta\mu$ as 
\begin{equation}
  \Delta^2 \approx(\delta\mu)^2
   -\left[\frac{(\mu-4\delta\mu/3)^3-(2\delta\mu)^3}
               {6(\mu-\delta\mu/3)^2}\right]^2.
   \label{neutgap}
\end{equation}
Here we have assumed $\mu_8=0$, which is known to be a reasonable 
approximation since nonzero $\mu_8$ can arise solely from nonzero $\Delta$.  
By taking note of the condition that Eqs.\ (\ref{gapg2SC}) and (\ref{neutgap}) 
are compatible, we can find that the gapless state is energetically 
favored over the normal and 2SC states at intermediate coupling \cite{SH}.
In fact, the condition for the presence of the
gapless and neutral solution approximately reads
\begin{equation}
 0.11\mu\lesssim\delta\mu\lesssim\frac{3}{10}\mu.
 \label{g2SCneut}
\end{equation}
(See Fig.\ 1 for an example of the gapless and neutral solution.)
By combining this relation with the condition, 
$\Delta_{\rm 2SC}/2\lesssim\delta\mu\lesssim\Delta_{\rm 2SC}$, 
for the presence of the gapless 2SC solution, we obtain
\begin{equation}
   0.22\mu \lesssim \Delta_{\rm 2SC} \lesssim \frac{3}{10}\mu. 
\end{equation}
This relation implies that above (or below) this region of $\Delta_{\rm 2SC}$,
the neutral solution, that is, the intersection of the neutral
constraint as shown by the dotted line in Fig.~1 and the solution of
the gap equation as shown by the solid line (or the line of $\Delta=0$) 
in Fig.~1, lies only in the region of the 2SC (or normal) phase \cite{SH}.

\begin{figure}[t]
\begin{center}
\includegraphics[width=10cm]{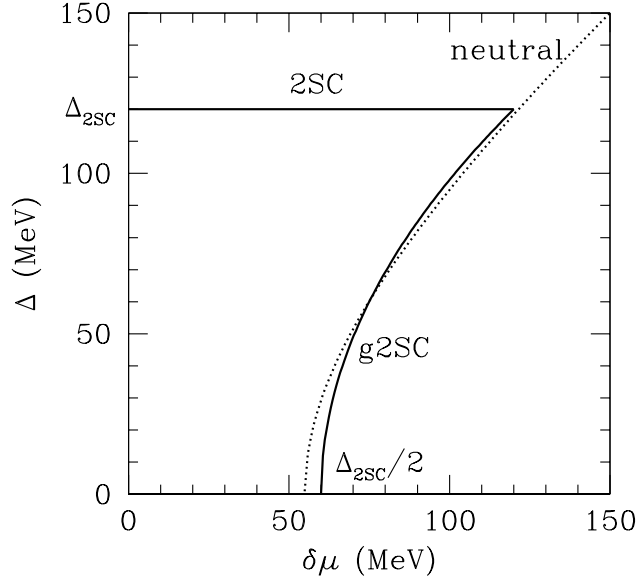}
\end{center}
\vspace{-0.5cm}
\caption{\label{fig1}
The homogeneous gap $\Delta$ as a function of $\delta\mu$ 
obtained for $\mu=500$ MeV and $\Delta_{\rm 2SC}(\delta\mu=0)=120$ MeV.  The 
upper solid line is the 2SC solution, and the lower solid line is the gapless 
2SC (g2SC) solution [see Eq.\ (\ref{gapg2SC})], while the dotted line
denotes the neutrality constraint [see\ Eq.\ (\ref{neutgap})].
The intersection of the g2SC solution and the dotted line corresponds to the
homogeneous neutral solution to the gap equation.  Note that the real 
2SC solution is dependent on $\delta\mu$ through the $\delta\mu$
dependence of $\bar\mu$, although such a dependence is negligible
under the approximation utilized here \cite{SH}.
}
\end{figure}

\section{Instability with respect to inhomogeneities}
\label{sec:inst}

     As mentioned in the previous section, the gapless 2SC state is stable 
against small homogeneous change in the gap magnitude.  However, it is
necessary to examine the stability against inhomogeneities.  In fact, it is 
known that the gapless 2SC state is unstable with respect to spontaneous 
generation of gauge fields \cite{HS2}.  In this section, we construct density 
functional theory to analyze the stability against inhomogeneities, and then 
describe typical instabilities which might occur in 
the gapless 2SC state and even in the 2SC state.

\subsection{Density functional theory}
\label{subsec:DFT} 

     We proceed to write the energy density functional allowing for 
infinitesimal inhomogeneities in the phases and magnitude of the order 
parameter and in the quark and electron densities, i.e., the position 
dependent densities $n_i({\bf r})\equiv n_i+\delta n_i({\bf r})$, gap 
magnitude $\Delta({\bf r})\equiv \Delta+\delta \Delta({\bf r})$, 
electromagnetic field ${\bf A}({\bf r})$, and gluon
fields ${\bf A}^\alpha ({\bf r})$ (as we shall see below, any nonzero 
gauge fields including constant ones correspond to inhomogeneities in 
the phases of the order parameter).  For 
inhomogeneities of spatial scale larger than the coherence length,
it is sufficient to adopt the local density approximation for
the bulk energy and keep the energies arising from the gap and density 
gradients up to leading order.  Thus, the total thermodynamic 
potential for given $\mu$ reads
\begin{equation}
 \Omega_{\rm tot}=
\int d^3 r\; \Omega_{\rm (g)2SC}(n_i({\bf r}),\Delta({\bf r}))
+\int d^3 r\; \Omega_g (n_i({\bf r}),\Delta({\bf r}),{\bf A}({\bf r}),
          {\bf A}^\alpha ({\bf r}))+E_C+\sum_\alpha E_C^\alpha.
\label{edf}
\end{equation}
Here $\Omega_{\rm 2SC}$ and $\Omega_{\rm g2SC}$ are given by Eqs.\ 
(\ref{omega2SC}) and (\ref{omegag2SC}), 
\begin{eqnarray}
 \Omega_g &=&\frac12 \sum_{fhab} B_{fa,hb}(\nabla n_{fa})\cdot (\nabla n_{hb})
    \nonumber \\ & &
   +2K_T^{(0)}\sum_{ab\alpha}
  \left|\left(\delta_{ab}\nabla
        -ig\frac{\lambda^{\alpha}_{ab}}{2}{\bf A}^\alpha
        +i\delta_{ab}\frac{e}{3}{\bf A}\right)\delta_{bB}\Delta\right|^2
    \nonumber \\ & &
   +2K_T^{(1)} \sum_{ab\alpha}
  \left|\delta_{aB}\Delta
       \left(\delta_{ab}\nabla-ig\frac{\lambda^{\alpha}_{ab}}{2}
        {\bf A}^\alpha
       +i\delta_{ab}\frac{e}{3}{\bf A}\right)\delta_{bB}\Delta\right|^2 
\label{grad}
\end{eqnarray}
is the gradient energy,
\begin{equation}
  E_C=\frac{1}{8\pi}  
      \int d^3 r_1 \int d^3 r_2 
             \frac{\rho({\bf r}_1)\rho({\bf r}_2)}{|{\bf r}_1-{\bf r}_2|}
\label{EC}                            
\end{equation}
with
\begin{equation}
  \rho=e\left(\sum_{fa} Q_f n_{fa} -n_e\right)
\end{equation}
is the electrostatic Coulomb energy, and
\begin{equation}
 \sum_\alpha E_C^\alpha=\frac{1}{8\pi}  
     \sum_\alpha \int d^3 r_1 \int d^3 r_2 
             \frac{\rho^\alpha({\bf r}_1)\rho^\alpha({\bf r}_2)}
                  {|{\bf r}_1-{\bf r}_2|}
\label{ECC}                            
\end{equation}
with
\begin{equation}
  \rho^{\alpha}=\frac12 g \sum_{fab} 
     \langle\bar{\psi}_{fa}\gamma^0 \lambda^{\alpha}_{ab}\psi_{fb}\rangle
\end{equation}
is the color Coulomb energy.  Here we have ignored the gluon contribution to 
the color density since it is of higher order in ${\bf A}^\alpha ({\bf r})$.  
For the covariant derivative, we have taken the form relevant for the gap
which has a $B$ direction in anti-color space, 
and $e/3$ corresponds to the electric
charge carried by a Cooper pair.  The local thermodynamic potential, up to 
second order in various inhomogeneities, can then be written as
\begin{eqnarray}
 \Omega_{\rm tot} &=& \Omega_0
\nonumber \\ & &
+\frac12 \int d^3 r \left[\sum_{i,j}
\frac{\partial^2 \Omega_{\rm (g)2SC}}
{\partial n_i \partial n_j}\delta n_i({\bf r}) \delta n_j({\bf r})
+2\sum_i \frac{\partial^2 \Omega_{\rm (g)2SC}}
{\partial n_i \partial \Delta}\delta n_i({\bf r}) \delta \Delta({\bf r})
+\frac{\partial^2 \Omega_{\rm (g)2SC}}
{\partial \Delta^2} (\delta \Delta({\bf r}))^2
\right]
\nonumber \\ & &
+\frac12 \sum_{abfh} \int \frac{d^3 q}{(2\pi)^3} q^2
B_{fa,hb} \delta n_{fa}({\bf q})\delta n_{hb}^*({\bf q})
+2(K_T^{(0)}+\Delta^2 K_T^{(1)})
\int \frac{d^3 q}{(2\pi)^3} q^2 |\delta \Delta({\bf q})|^2
\nonumber \\ & &
+\frac12 \sum_{\alpha\beta}\int d^3 r (m_M^{\alpha\beta})^2 
                         {\bf A}^\alpha ({\bf r}) {\bf A}^\beta ({\bf r})
+\frac12 \sum_\alpha \int d^3 r (m_M^\alpha)^2
                           {\bf A}^\alpha ({\bf r}) {\bf A}({\bf r})
+\frac12 \int d^3 r (m_M)^2 ({\bf A}({\bf r}))^2
\nonumber \\ & &
+\frac{1}{2} \int \frac{d^3 q}{(2\pi)^3} \frac{1}{q^2}
     |\rho({\bf q})|^2
+\frac{1}{2} \sum_\alpha \int \frac{d^3 q}{(2\pi)^3} \frac{1}{q^2}
     |\rho^\alpha ({\bf q})|^2,
  \label{local}
\end{eqnarray}
where $\Omega_0$ is the thermodynamic potential of the homogeneous state,
$\delta n_i ({\bf q})$, $\delta\Delta({\bf q})$, $\rho({\bf q})$, and
$\rho^\alpha ({\bf q})$ are the Fourier transforms of the corresponding 
inhomogeneities, and 
\begin{equation}
(m_M^{\alpha\beta})^2=\delta_{\alpha\beta}
 \left\{
  \begin{array}{ll}
    0,   &  \mbox{$\alpha=1$--3}  \\
    K_T^{(0)} g^2 \Delta^2,   &   \mbox{$\alpha=4$--7} \\
    \frac43 (K_T^{(0)}+\Delta^2 K_T^{(1)})
            g^2 \Delta^2,  &   \mbox{$\alpha=8$},
  \end{array}
 \right.
 \label{mm1}
\end{equation}
\begin{equation}
  (m_M^\alpha)^2 = \delta_{\alpha 8}\frac{4}{3\sqrt3}  
   (K_T^{(0)}+\Delta^2 K_T^{(1)}) ge \Delta^2,
 \label{mm2}
\end{equation}
and 
\begin{equation}
  (m_M)^2 = \frac{4}{9}  
   (K_T^{(0)}+\Delta^2 K_T^{(1)}) e^2 \Delta^2
 \label{mm3}
\end{equation}
are the Meissner masses squared for gluon-gluon, gluon-photon, and 
photon-photon channels, respectively.  Here we note that the first order terms
vanish due to equilibrium, i.e., 
\begin{equation}
 \frac{\partial\Omega_{\rm (g)2SC}}{\partial n_i} 
=\frac{\partial\Omega_{\rm (g)2SC}}{\partial \Delta}=0,
\end{equation}
and that the terms including both ${\bf A}$ or ${\bf A}^\alpha$ and 
$\delta \Delta$ or $\delta n_i$ do not appear up to second order.  We also 
neglect the electron contribution to the gradient energy, which is expected to
be small compared with the quark contribution that will be discussed below.

     We can simplify the color Coulomb energy by only retaining the components
of $\alpha=3,8$.  In fact, for $\alpha\neq3,8$, $\rho^\alpha$ turns out to be 
zero; any deviation of $\rho^\alpha$ from zero for $\alpha\neq3,8$ would
result in energy increase because there is no dependence on $\rho^\alpha$ in 
the other energy terms. (Vanishing $\rho^\alpha$ conforms to the color 
neutrality $\int d^3 r \rho^\alpha ({\bf r})=0$.)  Consequently, the color
Coulomb energy can be written as the sum of $E_C^{\alpha=3}$ and 
$E_C^{\alpha=8}$ given by
\begin{equation}
  E_C^{\alpha=3,8}=\frac{1}{2}  
      \int \frac{d^3 q}{(2\pi)^3} 
             \left(\frac{\sqrt3 g}{2}\right)^2 \frac{1}{q^2}
            \left|\sum_a Q^{\alpha=3,8}_a \delta n_a ({\bf q})\right|^2,
   \label{ECC2}
\end{equation}
where $\delta n_a=\sum_f \delta n_{fa}$ and $Q^{\alpha=3}_a 
=(1/\sqrt3,-1/\sqrt3,0)$.

     For later use, we write the expressions for the second derivatives
of the thermodynamic potential density as
\begin{equation}
\frac{\partial^2 \Omega_{\rm (g)2SC}}{\partial \Delta^2} 
=\frac{4{\bar\mu}^2}{\pi^2}
 \left[1-\theta(\delta\mu-\Delta)
  \left(\frac{\delta\mu}{\sqrt{(\delta\mu)^2-\Delta^2}}
     +\frac{\Delta^2}{{\bar\mu}^2}
  \ln\frac{\sqrt{(\delta\mu)^2-\Delta^2}+\delta\mu}{\Delta}\right)\right],
  \label{omegadd}
\end{equation}
\begin{equation}
\frac{\partial^2 \Omega_{\rm (g)2SC}}{\partial n_i \partial\Delta}
 =\frac{\partial \mu_i}{\partial\Delta}
 =
  \cases{
   \Delta\left[-\displaystyle{\frac{2}{\bar\mu}}
                  \ln\displaystyle{\frac{2\Lambda}{\Delta}}
  +\theta(\delta\mu-\Delta)\left(\displaystyle{\frac{2}{\bar\mu}}
   \ln\displaystyle{\frac{\sqrt{(\delta\mu)^2-\Delta^2}+\delta\mu}{\Delta}}
   -\displaystyle{\frac{1}{\delta\mu}} \right)
          \right],   
                          &   \mbox{$i=uR,uG$}  \cr 
   \Delta\left[-\displaystyle{\frac{2}{\bar\mu}}
                  \ln\displaystyle{\frac{2\Lambda}{\Delta}}
  +\theta(\delta\mu-\Delta)\left(\displaystyle{\frac{2}{\bar\mu}}
   \ln\displaystyle{\frac{\sqrt{(\delta\mu)^2-\Delta^2}+\delta\mu}{\Delta}}
   +\displaystyle{\frac{1}{\delta\mu}} \right)
          \right],   
                        &   \mbox{$i=dR,dG$} \cr 
    0,   &  \mbox{$i=uB,dB,e$},  \cr}
  \label{omegand}
\end{equation}
\begin{equation}
\frac{\partial^2 \Omega_{\rm (g)2SC}}{\partial n_i \partial n_j}
=\frac{\partial \mu_i}{\partial n_j}
=
 \cases{
   \displaystyle{\frac{\pi^2}{2{\bar\mu}^2}}
    \left[1+\left(\theta(\delta\mu-\Delta)
    \displaystyle{\frac{\delta\mu}{\sqrt{(\delta\mu)^2-\Delta^2}}}
                   \right)^{-1}\right],
                          &   \mbox{for $(i,j)=(ua,ub),(da,db)$} \cr
   \displaystyle{\frac{\pi^2}{2{\bar\mu}^2}}
    \left[1-\left(\theta(\delta\mu-\Delta)
   \displaystyle{\frac{\delta\mu}{\sqrt{(\delta\mu)^2-\Delta^2}}}
                   \right)^{-1}\right],
                          &   \mbox{for $(i,j)=(ua,db),(da,ub)$} \cr
    \delta_{ij}\left(\displaystyle{\frac{\pi}{3n_i}}\right)^{2/3}, 
                          &  \mbox{for $i=uB,dB,e$} \cr
                 0, & \mbox{for others},  \cr}
  \label{omegann}
\end{equation}
where $a=R,G$ and $b=R,G$.  In Eqs.\ (\ref{omegand}) and (\ref{omegann}) we 
have ignored higher order terms with respect to $\Delta/\Lambda$ and 
$\Delta/{\bar\mu}$.  In evaluating $\partial\mu_i/\partial\Delta$
we made use of $\partial n_i/\partial \Delta|_{n_i}=0$ in Eqs.\ (\ref{ni2SC})
and (\ref{nig2SC}).  Note that in the 2SC state, $\partial\mu_i/\partial n_j$ 
diverge for the combinations of $(i,j)$ participating in Cooper pairing.  This
feature, which comes from the fact that the corresponding part of the
thermodynamic potential (\ref{omega2SC}) does not depend explicitly on
$\delta\mu$, suggests that the number densities for $i=uR,uG,dR,dG$
would not change in the 2SC state.

     We now specify the parameters $B_{fa,hb}$, $K_T^{(0)}$, and $K_T^{(1)}$ 
characterizing the gradient energy.  To a first approximation, the density 
gradient term can be estimated as the Weizs{\" a}cker correction term $E_W$ 
\cite{Weiz}, which is the leading-order quantum correction to the Thomas-Fermi
model for the energy of an inhomogeneous ideal Fermi gas.  By following an 
argument for the nonrelativistic case \cite{Weiz}, it is straightforward to 
obtain the Weizs{\" a}cker term for an ultrarelativistic system as considered 
here as \cite{RTF}
\begin{equation}
E_W=\sum_{fa} \int d^3 r \frac{1}{72 (3 \pi^2)^{1/3}}
         \left(1+2\ln\frac{2\mu_{fa}}{m_f}\right)
         \frac{|\nabla n_{fa}|^2}{n_{fa}^{4/3}},
    \label{Wei}
\end{equation}
where we have introduced nonzero quark masses $m_f$ of order 5 MeV
in order to obtain a finite result.
The corresponding values of $B_{fa,hb}$ can be estimated as
$B_{fa,hb}=\delta_{ab}\delta_{fh} B^W_{fa}$ with 
$B^W_{fa}\sim 50(0.5~{\rm fm}^{-3}/n_{fa})^{4/3}$ MeV fm$^5$.  However, 
effects of the interaction between quarks can significantly modify the 
Weizs{\" a}cker term.  This is expected from the extended Thomas-Fermi model 
for atomic nuclei in which the interaction effects induce the terms 
including the difference and the sum of the proton and neutron density 
gradients and to increase the overall magnitude of the gradient term by a 
factor of 3--4.  Considering uncertainties due to these interaction effects, 
we take 
\begin{equation}
B_{fa,hb}=\delta_{ab}\delta_{fh} B^W_{fa}+B_S+B_A(3\delta_{ab}-1),
\end{equation}
with unknown parameters, $B^W_{fa}$, $B_S$, and $B_A$.  (Even
the Weizs{\" a}cker correction term can be modified by the interaction 
effects in a relativistic system \cite{RTF}.)  This choice leads to
\begin{eqnarray}
\frac12 \sum_{fhab} B_{fa,hb}(\nabla n_{fa})\cdot (\nabla n_{hb})
&=&\frac12 \sum_{fa} B^W_{fa} |\nabla \delta n_{fa}|^2
 +\frac12 B_S \left|\sum_a \nabla \delta n_a \right|^2 
\nonumber \\ & &
 + \frac12 B_A \left(|\nabla\delta n_R-\nabla\delta n_G|^2
                   +|\nabla\delta n_G-\nabla\delta n_B|^2
                   +|\nabla\delta n_B-\nabla\delta n_R|^2 \right).
\end{eqnarray}
Note that the present choice of $B_{fa,hb}$ allows for the different
interaction strengths for the color symmetric and antisymmetric combinations 
between $a$ and $b$, as expected from the gluon exchange interactions.

     For $K_T^{(0)}$ and $K_T^{(1)}$, we use the known weak coupling 
expressions as functions of $\Delta$, ${\bar\mu}$, and $\delta\mu$ \cite{RR}.
These expressions read
\begin{equation}
 K_T^{(0)}=\frac{{\bar\mu}^2}{6\pi^2 \Delta^2}
    \left[\left(1-2\frac{(\delta \mu)^2}{\Delta^2}\right)
          +\frac{2\delta\mu\sqrt{(\delta\mu)^2-\Delta^2}}{\Delta^2}
          \theta(\delta\mu-\Delta)\right],
    \label{KT0}
\end{equation}
and 
\begin{equation}
K_T^{(1)}=-\frac{{\bar\mu}^2}{12\pi^2 \Delta^4}
    \left[\left(1-4\frac{(\delta \mu)^2}{\Delta^2}\right)
          +\left(\frac{4\delta\mu\sqrt{(\delta\mu)^2-\Delta^2}}{\Delta^2}
                +\frac{\delta\mu}{\sqrt{(\delta\mu)^2-\Delta^2}}\right)
          \theta(\delta\mu-\Delta)\right].
    \label{KT1}
\end{equation}
Combining Eqs.\ (\ref{KT0}) and (\ref{KT1}) with Eqs.\ 
(\ref{mm1})--(\ref{mm3}), one can reproduce the Meissner masses in the weak 
coupling limit \cite{HS2}.  It is interesting to note that near the transition 
temperature, where the gap magnitude is suppressed, the stiffness parameter is
still similar to expression (\ref{KT0}) in magnitude, while the term 
associated with $K_T^{(1)}$ becomes negligible as compared with the term
associated with $K_T^{(0)}$.

\subsection{Chromomagnetic instability}
\label{sec:chromo}

     From the energy density functional (\ref{local}), we now proceed to 
consider possible instabilities in the 2SC and gapless 2SC phases.  We 
first examine chromomagnetic instabilities predicted to occur in both
phases \cite{HS2}.  The chromomagnetic instability is an
instability associated with the phase inhomogeneities or, equivalently,
${\bf A}^\alpha ({\bf r})$ and ${\bf A}({\bf r})$.  We set the other 
inhomogeneities to be zero.  Then, the sign of the Meissner 
masses squared or, equivalently, the sign of the specific combination of the 
stiffness parameters as in Eqs.\ (\ref{mm1})--(\ref{mm3}), plays a role in
determining the occurrence of the instability.  We thus have only to 
take note of the stiffness parameters, which is similar to the way 
Giannakis and Ren \cite{GR} approached the chromomagnetic instability 
associated with the $U(1)$ phase. 

     The instability related to negative $(m_M^{88})^2$, $(m_M^8)^2$, and 
$(m_M)^2$ occurs only in the gapless 2SC phase $(\delta\mu > \Delta)$, as 
can be seen from the expressions for $K_T^{(0)}$ and $K_T^{(1)}$.  This 
instability can be viewed as the tendency towards a plane-wave LOFF state 
in which the gap spatially oscillates like $e^{i{\bf q}\cdot{\bf r}}$
(phase oscillations).  This is because the $\alpha=8$ component of the 
$SU(3)$ phase acts like a usual $U(1)$ phase for the $RG$ pairing 
considered here for the 2SC state.

     In order to look further into the relation between the $SU(3)$ and $U(1)$ 
phases, we explicitly write down 
the supercurrents associated with the gradient of the $U(1)$ phase $\phi$ 
and the color $SU(3)$ phases $\phi^{\alpha}$.  A degenerate order 
parameter set of the 2SC states can be obtained by transforming the 2SC state 
specified in the previous section under global $U(1)$ and color rotation as
\begin{equation}
   d_a =\left(e^{i\phi}e^{i\sum_\alpha \frac{\lambda^\alpha}{2}
        \phi_\alpha}\right)_{aB}\Delta.
   \label{op}
\end{equation}
By inserting this order parameter into $\delta_{aB}\Delta$ in
the gradient energy (\ref{grad})
one can obtain the supercurrents from \cite{II}
\begin{equation}
 J_i= \frac{\partial \Omega_g}{\partial A_i},~~~
 J_i^\alpha=\frac{\partial \Omega_g}{\partial A_i^\alpha}.
    \label{currents}  
\end{equation}
The results read
\begin{equation}
 J^\alpha_i=\left\{
  \begin{array}{ll}
             0,            &   \mbox{$\alpha=1$--3} \\
  K_T^{(0)}g\Delta^2 \partial_i \phi_\alpha -(m_M^{\alpha\alpha})^2 
      A^\alpha_i,
                           &   \mbox{$\alpha=4$--7} \\
 -(K_T^{(0)}+\Delta^2 K_T^{(1)})g\Delta^2
  \left(\displaystyle{\frac{4}{\sqrt3}} \partial_i \phi
       -\displaystyle{\frac{4}{3}} \partial_i \phi_8 \right)
      -(m_M^{88})^2 A^8_i -(m_M^8)^2 A_i,
                           &  \mbox{$\alpha=8$},  
  \end{array}
 \right. 
\end{equation}
and
\begin{equation}
  J_i=\frac{e}{\sqrt3 g}J_i^8.
\end{equation}
It is thus evident that the gauge fields play the role of the phase gradients 
in the sense that any supercurrent pattern created by the phase gradients 
can be reproduced by the gauge fields alone.

     From these supercurrents, we can find that the system, if undergoing the 
chromomagnetic instabilities associated with ${\bf A}$ and/or ${\bf A}^8$,
would tend to a plane-wave LOFF state.  In fact, the proportionality of 
${\bf J}$ to ${\bf J}^8$ implies that any inhomogeneous state created by
${\bf A}^8$ could be created by ${\bf A}$.  It is thus sufficient to consider
spontaneous generation of ${\bf A}$.  As long as ${\bf A}$ is small and 
uniform, the resultant supercurrent can be reproduced by 
$\phi({\bf r})={\bf q}\cdot{\bf r}$ with ${\bf q}=\frac{e}{3}{\bf A}$, 
which in turn characterizes a plane wave LOFF state.

     There is another type of chromomagnetic instability, which is associated 
with negative $(m_M^{44})^2$, $(m_M^{55})^2$, $(m_M^{66})^2$, and 
$(m_M^{77})^2$.  This instability appears 
not only in the gapless 2SC state but also in a part of the 2SC state that 
fulfills $\delta\mu > \Delta/\sqrt2$.  This instability couples with the phase
gradients of $\phi_4$--$\phi_7$.  For small and uniform ${\bf A}^\alpha$ with 
$\alpha=4$--7, the state can be described by 
$\phi^\alpha ({\bf r})={\bf q}\cdot{\bf r}$ with ${\bf q}=-g{\bf A}^\alpha$.
Consequently, this state corresponds to a LOFF state characterized by 
the color rotation $\exp(i\lambda^\alpha {\bf q}\cdot{\bf r}/2)$ of the
gap.  Note, however, that in the gapless 2SC state, the absolute values of 
$(m_M^{44})^2, \cdots, (m_M^{77})^2$ are not larger than
those of $(m_M^{88})^2$, $(m_M^8)^2$, and $(m_M)^2$ [see 
Eqs.\ (\ref{mm1})--(\ref{mm3}) with Eqs.\ (\ref{KT0}) and (\ref{KT1})]. 
We can thus conclude that in the gapless 2SC state, the tendency to the LOFF 
state with color rotations is weaker than that with $U(1)$ rotations.

    Note that the instabilities considered here do not involve any 
variation of the gap amplitude or the particle densities.  We will thus 
consider the cases in which such variations occur in the next two 
subsections and discuss the possibility of instabilities with respect to
them.

\subsection{Gap amplitude instability}
\label{sec:gai}

    We now consider inhomogeneities in the gap amplitude by setting
the other inhomogeneities to be zero.  We can see such an
instability from the second derivative of $\Omega_{\rm (g)2SC}$ with
respect to $\delta\Delta({\bf q})$.  Note that the second derivative 
(\ref{omegadd}), which is relevant for ${\bf q}=0$, can be negative only 
in the gapless SC state, implying that this comes from the presence 
of the gapless modes.  The associated instability remains in the limit 
of ${\bf q}\to0$ in gapless neutral superfluids \cite{Sarma} which are 
generally unstable with respect to homogeneous (${\bf q}=0$) change in the gap 
amplitude, while neutrality constraints keep the gapless charged 
superconductors stable against homogeneous change in the gap magnitude.  
This is because this change is accompanied by homogeneous change in 
the quark and electron densities, leading to increase in the total 
kinetic energy.

    It is important to note that at nonzero ${\bf q}$, the system becomes
more unstable by having the energy lowered by the gradient
energy term associated with inhomogeneities in the gap amplitude
in the gapless phase where the stiffness parameters are negative. 
Then the neutrality constraints no longer hinder the system from
undergoing the instability at nonzero ${\bf q}$.  This implies the 
tendency towards a LOFF state in which the gap amplitude oscillates.  
In general, the gap amplitude instability develops in the 
gapless state together with the chromomagnetic instability,
though we will refer to this situation simply as the instability with 
respect to the gap amplitude.

\subsection{Stability against density fluctuations}
\label{sec:dci}

     We finally consider the case in which inhomogeneities are present only
in the quark and electron densities.  The possible occurrence of 
instabilities depends on the structure of 
$\partial^2 \Omega_{\rm (g)2SC}/\partial n_i \partial n_j$.  In order for the
system to be stable against small density modulations, all eigenvalues 
of this hermitian matrix must be positive, i.e., any minors of 
the determinant must be positive.  This condition holds both for the 2SC state
and for the gapless 2SC state.

     It is nonetheless important to note that the Fermi surface separation
between $d$ and $u$ quarks can vary nonuniformly in the gapless 2SC state, 
which is driven by the gap amplitude instability discussed above in such a 
way that as the gap amplitude increases, the Fermi surface separation 
decreases.  Generally, such density variation produces a positive gradient 
energy.  If this energy dominates over the gradient energy associated with 
inhomogeneities in the gap amplitude, one can expect that
the system tends to a coexisitng phase of large BCS (2SC) domains with 
small Fermi surface separation and large normal domains with large separation
\cite{RR}.  This phase can be viewed as a state in which clustering of $d$ 
quarks occurs within normal domains.

     In order to clarify what kind of structure this clustering takes on, one 
needs to take full account of inhomogeneities in the densities and the gap 
amplitude; this account would automatically involve charge screening, which 
plays an important role in determining the structure.  This structure lies in 
between the strong screening limit (phase separation) and the screeningless 
limit, as will be quantified in the next section.  We remark that the same 
kind of phase separation was recently observed in a superfluid gas of atomic 
fermions with unequal numbers of two components \cite{Part}.

\section{Effective potential for the gap magnitude}
\label{sec:pot}

     In this section we systematically consider inhomogeneities in the quark 
and electron densities and in the phases and amplitude of the order parameter
to determine the kind of inhomogeneities that spontaneously develop in the 
2SC and gapless 2SC states.  In doing so, we first focus on the bulk 
(${\bf q}$ independent) part of
the effective potential for the variation of the gap amplitude.  
We then include the effect of the Coulomb and gradient parts in the effective 
potential and thereby clarify what state the system tends to.

     Let us begin with the expression for the effective potential $v(q)$
for given $\delta\Delta({\bf q})$, which can be obtained from the local
thermodynamic potential (\ref{local}) as
\begin{eqnarray}
 \Omega_{\rm tot} &=& \Omega_0
\nonumber \\ & &
+\frac12 \int \frac{d^3 q}{(2\pi)^3} v(q) |\delta \Delta({\bf q})|^2
\nonumber \\ & &
+\frac12 \sum_{\alpha\beta}\int d^3 r (m_M^{\alpha\beta})^2 
                       {\bf A}^\alpha ({\bf r}) {\bf A}^\beta ({\bf r})
+\frac12 \sum_\alpha \int d^3 r (m_M^\alpha)^2
                       {\bf A}^\alpha ({\bf r}) {\bf A}({\bf r})
+\frac12 \int d^3 r (m_M)^2 ({\bf A}({\bf r}))^2,
  \label{local2}
\end{eqnarray}
where
\begin{eqnarray}
  v(q)&=&v_0
      +4(K_T^{(0)}+\Delta^2 K_T^{(1)}) q^2 
      +\sum_{fhab} [B^W_{fa} \delta_{ab}\delta_{fh}+B_S+B_A(3\delta_{ab}-1)]
             \frac{\delta n_{fa}({\bf q})\delta n^*_{hb}({\bf q})}
             {|\delta \Delta({\bf q})|^2} q^2 
\nonumber \\ & &
      +\frac{e^2\left|\sum_{fa} Q_f \delta n_{fa}({\bf q})
            -\delta n_e({\bf q})\right|^2+
      (\sqrt3 g/2)^2\sum_{\alpha=3,8}\left|\sum_a Q^{\alpha}_a 
      \delta n_a ({\bf q})\right|^2}
     {|\delta \Delta({\bf q})|^2}\frac{1}{q^2},
  \label{vq}
\end{eqnarray}
with the bulk contribution
\begin{equation}
v_0=\frac{\partial^2 \Omega_{\rm (g)2SC}}{\partial \Delta^2}
      +\sum_i\left[\frac{\delta n_i({\bf q})}{\delta \Delta({\bf q})}
            +\frac{\delta n_i^*({\bf q})}{\delta \Delta^*({\bf q})}
                             \right]
    \frac{\partial^2 \Omega_{\rm (g)2SC}}{\partial \Delta \partial n_i}
      +\sum_{ij}\frac{\delta n_i({\bf q})\delta n^*_j({\bf q})}
             {|\delta \Delta({\bf q})|^2}
    \frac{\partial^2 \Omega_{\rm (g)2SC}}{\partial n_i \partial n_j}.
   \label{v0}
\end{equation}
From the condition that this effective potential takes on a minimal value,
i.e., $\partial \Omega_{\rm tot}/\partial \delta n_i({\bf q})=0$,
we acquire the relations
\begin{equation}
  \frac{\partial\mu_e}{\partial n_e}\delta n_e({\bf q})
  =\frac{e^2}{q^2} \left[\sum_{fa}Q_f \delta n_{fa}({\bf q})
                         -\delta n_e({\bf q})\right]
 \label{varne}
\end{equation}
and
\begin{eqnarray}
  \sum_i \frac{\partial\mu_i}{\partial n_{fa}}\delta n_i({\bf q})
  &=&-\frac{Q_f e^2}{q^2} \left[\sum_{hb} Q_h \delta n_{hb}({\bf q})
                          -\delta n_e({\bf q})\right]
   -\left(\frac{\sqrt3 g}{2q}\right)^2 \sum_{\alpha=3,8}
     Q_a^\alpha \sum_{hb} Q_b^\alpha \delta n_{hb}({\bf q})
 \nonumber \\ & & 
   -q^2 \sum_{hb} [B^W_{fa} \delta_{ab}\delta_{fh}+B_S+B_A(3\delta_{ab}-1)]
    \delta n_{hb}({\bf q}) 
   -\frac{\partial\mu_{fa}}{\partial\Delta}\delta \Delta({\bf q}).
 \label{varnq}
\end{eqnarray}

    In the following, to a first approximation, we search for the 
solution to Eqs.\ (\ref{varne}) and (\ref{varnq}) in the absence of finite 
size corrections, namely, in the absence of the terms coming from
the gradient and Coulomb energies.  This solution is good enough to control 
the sign of $v(q)$.  We then extend the calculations to the case with finite 
size corrections to clarify the detailed spatial structure of the 
instabilities of the system as function of $\delta\mu$.

\subsection{Case without finite size corrections}

    In Secs.\ \ref{sec:gai} and \ref{sec:dci}, we have found that the gapless
state is unstable with respect to the gap amplitude oscillations 
and stable with respect to density fluctuations 
by considering $\delta\Delta({\bf q})$ and 
$\delta n_i({\bf q})$ separately.  The bulk part, $v_0$, of the effective 
potential, which we will focus on in this subsection, is expected to 
clarify how $\delta n_i({\bf q})$ is related with $\delta\Delta({\bf q})$.

     Once one neglects the Coulomb and gradient parts of the effective 
potential, one is allowed by the symmetry of $v_0$ in color and flavor space 
to set
\begin{equation}
   \sum_{a=R,G}\delta n_{ua}\equiv \delta n_- +\delta n_+,~~~
   \sum_{a=R,G}\delta n_{da}\equiv -\delta n_- +\delta n_+.
\end{equation}
Then, one can obtain the relation 
between $\delta\Delta({\bf q})$ and $\delta n_i({\bf q})$ from Eqs.\ 
(\ref{varne}) and (\ref{varnq}), by retaining the term of leading order
in $\delta\mu/{\bar\mu}$ for $\partial\mu_i/\partial\Delta$, 
Eq.\ (\ref{omegand}), as
\begin{equation}
\delta n_- = \delta n_+ =0,\qquad
\delta n_e=\delta n_{uB}=\delta n_{dB}=0, 
\label{2SCsol}
\end{equation}
for the 2SC state, and as
\begin{equation}
\delta n_-
=\frac{{\bar\mu}^2 \delta\Delta}{\pi^2}
\frac{\Delta}{\sqrt{(\delta\mu)^2-\Delta^2}},\qquad
\delta n_+=0,
\label{dnu}
\end{equation}
\begin{equation}
\delta n_e=\delta n_{uB}=\delta n_{dB}=0, 
\label{g2SCsol}
\end{equation}
for the gapless 2SC state.  

     The neglect of the gradient and Coulomb energies is a good 
approximation as long as the typical scale of the 
incompressibilities $\partial \mu_i/\partial n_j$, which is of order
$\pi^2/\mu^2$, is sufficiently large.  However, it is important to note that
the color Coulomb energies can be comparable to or even greater than the 
bulk contribution.  It is thus reasonable to keep the color Coulomb energies 
vanishingly small.  Due to $\delta n_+=0$, $E_C^{\alpha=8}=0$, while,
from Eq.\ (\ref{ECC2}), $E_C^{\alpha=3}=0$ leads to
\begin{equation} 
  \delta n_{uR}+\delta n_{dR}=\delta n_{uG}+\delta n_{dG}.
   \label{EC30}
\end{equation}

     By substituting the resulting $\delta n_i$ into $v_0$, we obtain
\begin{eqnarray}
  v_0&=&\frac{4{\bar\mu}^2}{\pi^2}
  \left\{1-\theta(\delta\mu-\Delta)\left[
  \frac{\delta\mu}{\sqrt{(\delta\mu)^2-\Delta^2}}
   \left(1+\frac{\Delta^2}{2(\delta\mu)^2}\right)
   +\frac{\Delta^2}{{\bar\mu}^2}
\ln\frac{\sqrt{(\delta\mu)^2-\Delta^2}+\delta\mu}{\Delta}
   \right]\right\}.
  \label{v02}
\end{eqnarray}
The sign of $v_0$ predominantly determines the sign of $v(q)$, as
we will numerically confirm later.  Note that $v_0$ is positive (negative)
in the 2SC (gapless 2SC) state.  This implies
that the gapless 2SC state tends to be unstable with respect to small
variations of the gap amplitude, while the 2SC 
state being stable.  Consequently, the 2SC state is unstable only 
to the variation of the phases of $\alpha=4$--7; this instability 
occurs when $\delta\mu>\Delta/\sqrt2$ as shown in Sec.\ \ref{sec:chromo}.

     Note that the gap amplitude instability, which is predicted to occur 
in the gapless 2SC state, results in the $d$ quark clustering.
In fact, the density difference 
between $u$ and $d$ quarks becomes smaller with increasing 
$\delta\Delta({\bf q})$, as can be seen from Eq.\ (\ref{dnu}).

\subsection{Case with finite size corrections}

    We now proceed to take into account finite size corrections due to
the Coulomb and gradient energies in evaluating the effective potential.  
As we shall see, the sign of the
gradient contribution and the effect of electric charge screening
play a role in determining the configuration of inhomogeneities that develop
spontaneously.

    By substituting the bulk solutions (\ref{2SCsol})--(\ref{g2SCsol}) 
into Eq.\ (\ref{vq}), we obtain the gradient contribution
to the effective potential as
\begin{eqnarray}
 && v_g q^2 \equiv 4(K_T^{(0)}+\Delta^2 K_T^{(1)}) q^2 
      +\sum_{fhab} [B^W_{fa}\delta_{ab}\delta_{fh}+B_S+B_A(3\delta_{ab}-1)]
             \frac{\delta n_{fa}({\bf q})\delta n^*_{hb}({\bf q})}
             {|\delta \Delta({\bf q})|^2} q^2 
\nonumber \\ &&\quad
=\Biggl\{\frac{1}{3\pi^2}\frac{{\bar\mu}^2}{\Delta^2}
   \left[1-\theta(\delta\mu-\Delta)
   \frac{\delta\mu}{\sqrt{(\delta\mu)^2-\Delta^2}}\right]
 +\!\sum_{f=u,d,~a=R,G}\!\! B^W_{fa} \theta(\delta\mu-\Delta)
    \left[\frac{\Delta {\bar\mu}^2}
       {2\pi^2 \sqrt{(\delta\mu)^2-\Delta^2}}\right]^2\Biggr\}q^2.
    \label{vg}
\end{eqnarray}
Here, for the gapless 2SC state, we have used
\begin{equation}
  \delta n_{uR}=\delta n_{uG}=-\delta n_{dR}=-\delta n_{dG}
  =\frac12 \delta n_-.
\end{equation}
This relation can be obtained by combining Eq.\ (\ref{EC30}) with Eq.\ 
(\ref{dnu}) and by minimizing the Weizs{\"a}cker term under $B^W_{uR}=B^W_{uG}$
and $B^W_{dR}=B^W_{dG}$, which stem from the relations, $n_{uR}=n_{uG}$ and 
$n_{dR}=n_{dG}$, given by Eq.\ (\ref{nig2SC}) as well as the relations, 
$\mu_{uR}=\mu_{uG}$ and $\mu_{dR}=\mu_{dG}$, given by Eq.\ (\ref{mufa}).
We remark that the density gradient contributions associated with the 
parameters $B_S$ and $B_A$ vanish in the absence of $\delta n_{R,G,B}$.

     From $v_g$ we examine how instabilities in the gap magnitude develop
in the gapless 2SC state.  For ${\bar\mu}\sim500$~MeV, $\Delta\sim100$
MeV, and $B^W_{fa}\sim50$ MeV fm$^5$, $v_g$ is negative, due to the negative
gap gradient energy, except in the immediate vicinity of the gapless onset 
$\delta\mu=\Delta$, i.e., $v_g$ takes a positive value only for 
$\delta\mu/\Delta<x_2\sim 1.011$.  
When $v_g$ is negative, the system tends to a LOFF state with amplitude 
oscillations since a larger $q$ is more favorable.  In this LOFF state, 
the amplitude oscillation is allowed to occur at a spatial scale of the order 
of the coherence length.  Note that the local thermodynamic potential 
(\ref{local2}) suggests an even stronger instability of the system due to the 
variation of the gauge fields, which is independent of $\delta \Delta$ 
because the terms like $\partial^2 \Omega_{\rm (g)2SC}/\partial A_i^\alpha 
\partial \Delta$ vanish up to second order in the inhomogeneities.  Since this 
additional instability is particularly strong for ${\bf A}$ and/or ${\bf A}^8$ 
as characterized by the negatively divergent behavior of $(m_M^{88})^2$, 
$(m_M^8)^2$, and $(m_M)^2$ near the onset of the gapless 2SC state, the 
eventual LOFF state would presumably be described as a superposition of 
plane waves involving the appreciable change in the amplitude.  Note,
however, that this state still undergoes chromomagnetic instabilities 
associated with $\alpha=4$--7 \cite{Gorbar}.  In any case, we
shall refer to this state with $v_g<0$ (large $q^2$) as the LOFF 
state with amplitude oscillations.

     It is instructive to consider the case in which $v_g$ is positive 
although this is limited to the immediate vicinity of $\delta\mu=\Delta$
for the typical case in which ${\bar\mu}\sim500$~MeV, $\Delta\sim100$
MeV, and $B^W_{fa}\sim50$ MeV fm$^5$.
This situation implies the tendency to either a BCS-normal mixed state or 
phase separated state.  As we mentioned before, the mixed or phase 
separated state is not clearly separable from the LOFF state with amplitude 
oscillations.  Qualitatively, if $q^2$ is large as suggested by $v_g<0$ then 
the resultant inhomogeneity is characteristic of the amplitude LOFF 
state.  There comes out the mixed phase next as $q^2$ decreases and finally the
phase separated state for even smaller $q^2$.  Let us now quantify this
criterion by determining the characteristic scale of $q^2=Q^2$ that 
distinguishes between the mixed and phase separated states.  The typical scale 
of the spatial structure of the mixed phase, $2\pi Q^{-1}$, is determined in 
such a way as to minimize the sum of the gradient and Coulomb contributions to 
the effective potential.  This scale arises because the gradient energy
increases with $q$ while the Coulomb energy decreases with $q$.
In evaluating the Coulomb contribution, it is important to 
allow for the variation of the electron density, which acts to screen the 
charge distribution formed by the $R$ and $G$ quark components.  The 
Coulomb contribution including these screening corrections can be expressed as
\begin{eqnarray}
 v_C &\equiv& 
    \frac{\partial \mu_e}{\partial n_e}
    \frac{|\delta n_e ({\bf q})|^2}{|\delta \Delta({\bf q})|^2}
   +\frac{e^2 \left|\sum_{fa}Q_f\delta n_{fa}({\bf q})-\delta n_e({\bf q})
             \right|^2}{|\delta \Delta({\bf q})|^2 q^2}
 \nonumber \\ 
   &\simeq&\frac{e^2 |\delta n_{-}({\bf q})|^2}
             {(q^2+k_{\rm TF}^2)|\delta\Delta({\bf q})|^2},
   \label{vc}
\end{eqnarray}
where 
\begin{equation}
  k_{\rm TF}=\frac{e\mu_e}{\pi}
  \label{TF}
\end{equation}
is the inverse of the Thomas-Fermi screening length.  Here we have 
used Eq.\ (\ref{varne}) in evaluating $\delta n_e$, and we have ignored a 
possible screening by quarks, since it would lead to a nonvanishing net 
color charge which should be energetically disfavored.

     By combining Eqs.\ (\ref{vg}) and (\ref{vc}), we can estimate the 
value of $q=Q$ that minimizes the sum of $v_g q^2$ and $v_C$ as
\begin{equation}
Q^2=-k_{\rm TF}^2+\left(\frac{e^2}{\beta_g}\right)^{1/2},
  \label{Q}
\end{equation}
where $\beta_g\equiv v_g (|\delta\Delta|/|\delta n_{-}|)^2$.  At $q=Q$, the 
sum reads
\begin{equation}
   v_g Q^2+v_C=\left(2\sqrt{\beta_g e^2}-\beta_g k_{\rm TF}^2\right)
           \frac{|\delta n_{-}|^2}{|\delta\Delta|^2}.
\end{equation}
When the value of $\beta_g$ is 
sufficiently small that $Q^2\gtrsim k_{\rm TF}^2$, it is expected that the 
normal-BCS mixed state occurs in the form of a Coulomb lattice of 
periodicity of order $2\pi Q^{-1}$.  We remark that a similar situation occurs 
in the liquid-gas mixed phase of nuclear matter at subnuclear densities 
\cite{NM}.

     Near the gapless onset $\delta\mu/\Delta=1$ we can expand
$Q^2$ with respect to $\epsilon\equiv\delta\mu/\Delta-1$ as
\begin{equation}
 Q^2\approx -k_{\rm TF}^2+\frac{e}{\sqrt{B^W}}
  +\frac{\sqrt{2}\pi^2e\sqrt{\epsilon}}
  {6(B^W)^{3/2}\Delta^2 \bar{\mu}^2}
  +\frac{\pi^2 e(\pi^2-4B^W \Delta^2\bar{\mu}^2)\epsilon}{12(B^W)^{5/2}
  \Delta^4\bar{\mu}^4}+\cdots,
\label{eq:Q}
\end{equation}
where we denote $B_{fa}^W$ simply by $B^W$.  In the typical case in which 
$\Delta=100$ MeV and $B^W=B^W_0=50$~MeV~fm$^5$, near the gapless onset, we
obtain $Q^2\simeq e/\sqrt{B^W}\simeq 0.6\mbox{ fm}^{-2}\gg k_{\rm TF}^2\simeq
0.01\mbox{ fm}^{-2}$.  In this case, the normal-BCS mixed phase
is expected in the form of a Coulomb lattice of periodicity $\sim
2\pi/Q\simeq 8$~fm.  We plot $Q^2$ for this case in Fig.~2(a) and it is 
clear from the figure that $Q^2$ is much larger than $k_{\rm TF}^2$ in the 
whole region up to $\delta\mu/\Delta=x_2$.

       The value of $Q^2$, as seen from Eq.~(\ref{eq:Q}), depends strongly on 
the value of $B^W$.  For sufficiently large values of $B^W$, 
we obtain a negative 
value of $Q^2$.  This indicates that the real solution is $Q=0$ and that the 
electron screening is sufficiently strong to separate the system into the 
normal and BCS (2SC) regions that are locally charge neutral.  This situation 
is expected to continue until $Q^2$ amounts to $\sim k_{\rm TF}^2$.  Near
$Q=k_{\rm TF}$, at which we define $x_1=\delta\mu/\Delta$, the electron 
screening ceases to ensure the local charge neutrality.  
We have calculated $Q^2$ for the case of $B^W=30B^W_0$ and depicted the 
behavior of $Q^2$ in Fig.~2(b).  In this case, the decreasing behavior of 
$Q^2$ is dominated by the term, $-k_{\rm TF}^2$, in Eq.\ (\ref{eq:Q}).  
We also note that in this case $x_2$ does not exist since $v_g$ is
positive for $\delta\mu/\Delta>1$.  In fact, $v_g$ can be negative only when
\begin{equation}
     B^W < \frac{\pi^2}{6\Delta^2{\bar\mu}^2} \equiv B^W_c,
\label{Bcrit}
\end{equation}
where $B^W_c$ amounts to $\sim200$ MeV fm$^5$ for
$\Delta\sim100$ MeV and ${\bar\mu}\sim500$ MeV.

\begin{figure}
\begin{center}
 \begin{minipage}{7cm} {\large (a)}\\
  \includegraphics[width=7cm]{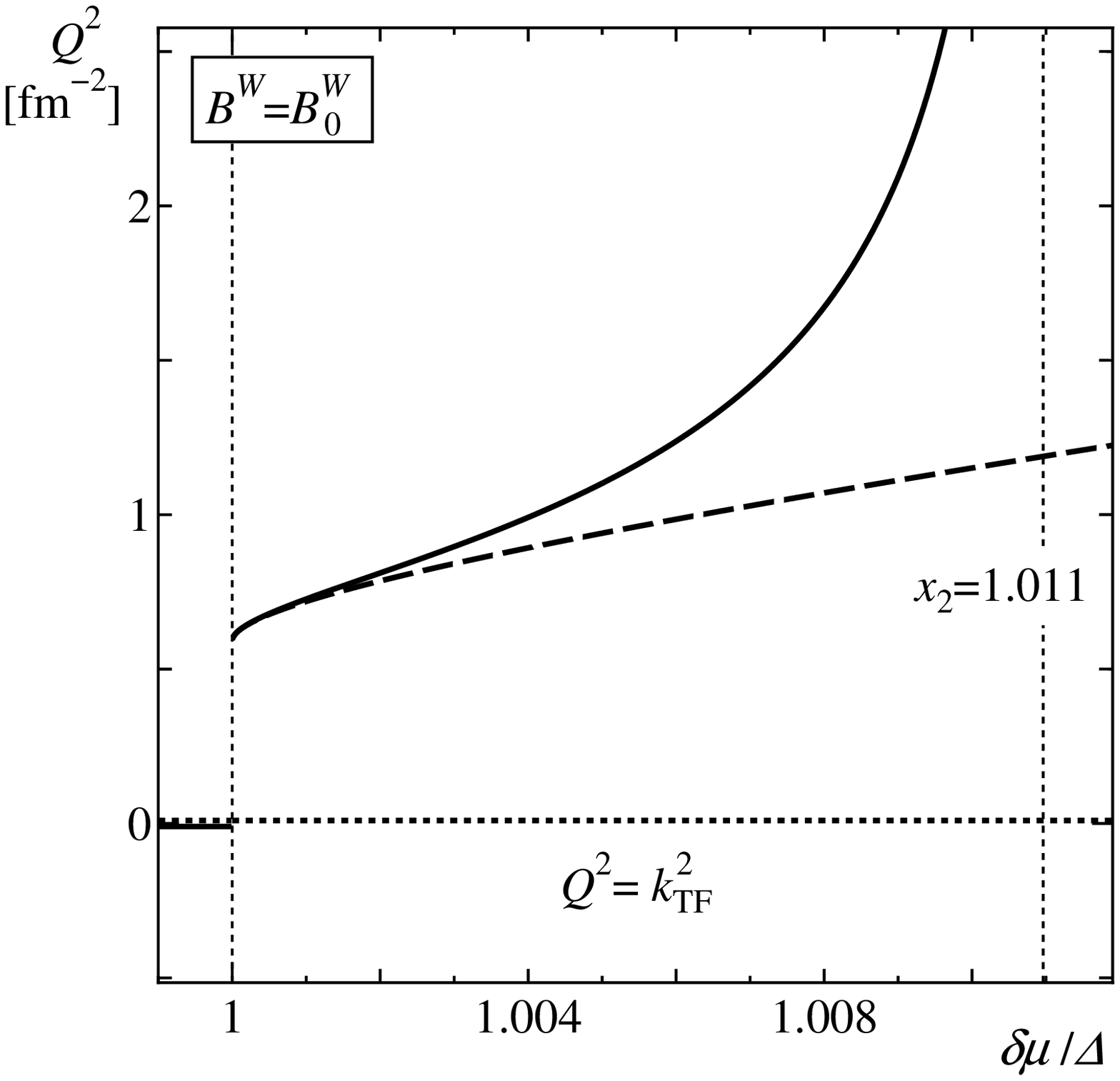}
 \end{minipage} \hspace{1cm}
 \begin{minipage}{7cm} {\large (b)}\\
  \includegraphics[width=7cm]{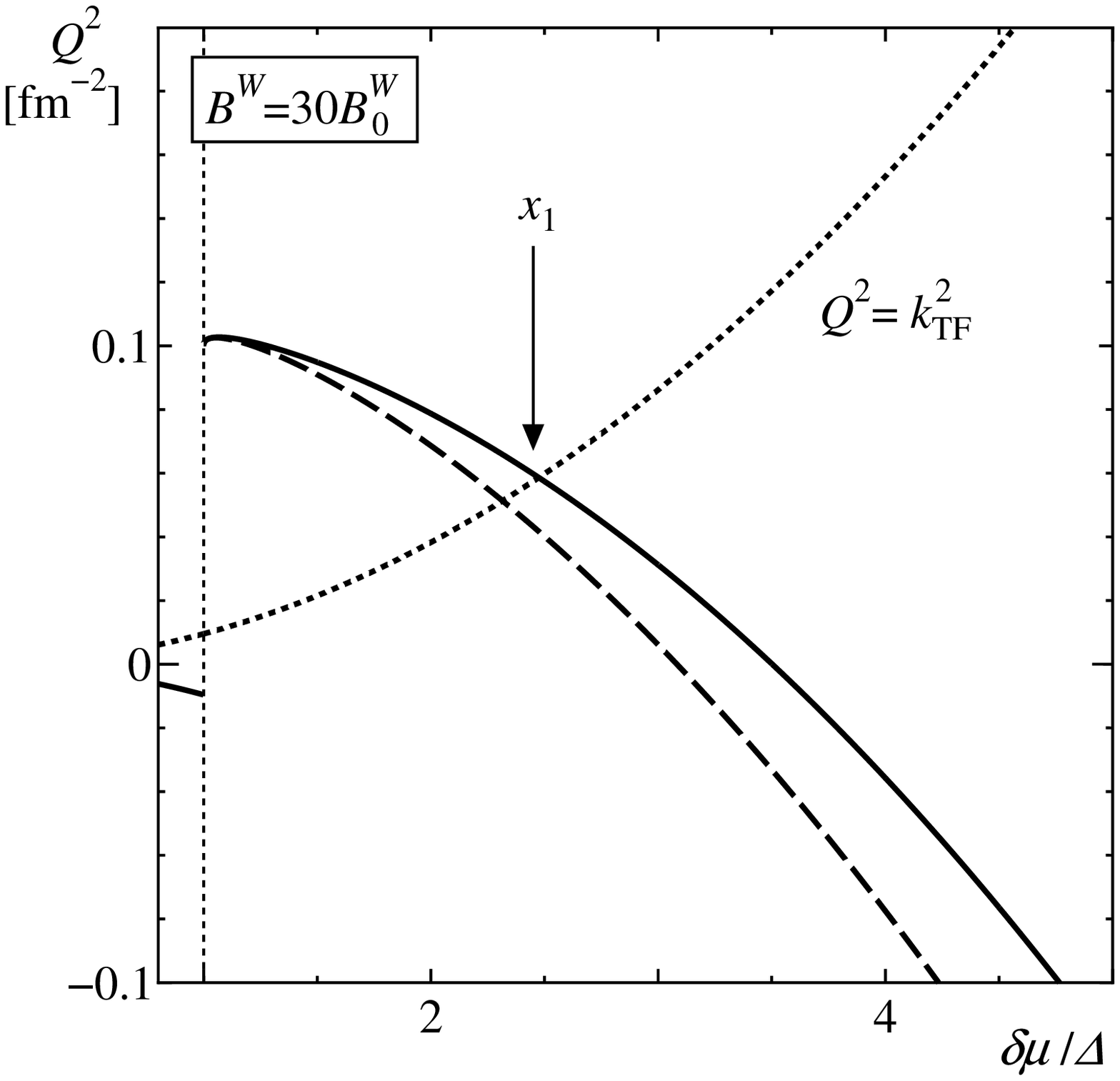}
 \end{minipage}
\end{center}
\caption{Spatial momentum squared $Q^2$ at which the system is
most unstable to the gap amplitude variation.  The solid lines are the 
results from Eq.\ (\ref{Q}), and the dashed lines are the results from
Eq.\ (\ref{eq:Q}).  (a) The typical case $\Delta=100\mbox{ MeV}$ and
$B^W=B^W_0=50\mbox{ MeV fm}^5$.  For $1<\delta\mu/\Delta<x_2=1.011$, 
$Q^2$ is much larger than $k_{\rm TF}^2$ indicating that the system 
takes a form of the Coulomb lattice with periodicity $\sim2\pi/Q$.  
(b) For large $B^W=30B^W_0$ there appears the region in which $Q^2$ is 
smaller than $k_{\rm TF}^2$.  Then the screening effect is so significant 
that the system can be phase separated.
}
\end{figure}

     Recall that the local thermodynamic potential (\ref{local2})  
suggests an even stronger instability due to the variation of the gauge 
fields.   This implies the possibility that the normal state in the mixed 
phase is replaced by the LOFF state without amplitude oscillations.
Near $\delta\mu=\Delta$, $|\delta n_-/\delta\Delta|$ is far larger
than the typical scale $\mu^2/\pi^2$ of the density of states, indicating that
the gap might remain in the region where the Fermi surface separation
between $u$ and $d$ quarks is maximal.  Consequently, the LOFF state, which 
occurs only for nonzero $\Delta$, could appear in such a region.

    In Fig.\ 3, we summarize the instabilities of the homogeneous system
with respect to various inhomogeneities and the resultant possible states on 
the $B^W$ versus $\delta\mu/\Delta$ plane.  Through the last term in the 
thermodynamic potential (\ref{omegag2SC}), the presence of the gapless modes
in the gapless 2SC state underlies all the instabilities but the 
chromomagnetic instability that occurs also in the 2SC state.  It is 
noteworthy that no LOFF state with amplitude oscillations is expected in the 
gapless state when $B^W$ is larger than the critical value $B^W_c$.
We emphasize that all we can know from the present stability analysis is how 
the system is driven by infinitesimal variations of the densities and the 
order parameter.
This property, while being suggestive of the final destination of the 
unstable system, does not tell us what the ground state really is 
(see Ref.\ \cite{Shov} for its candidates).

\begin{figure}
\begin{center}
\includegraphics[width=10cm]{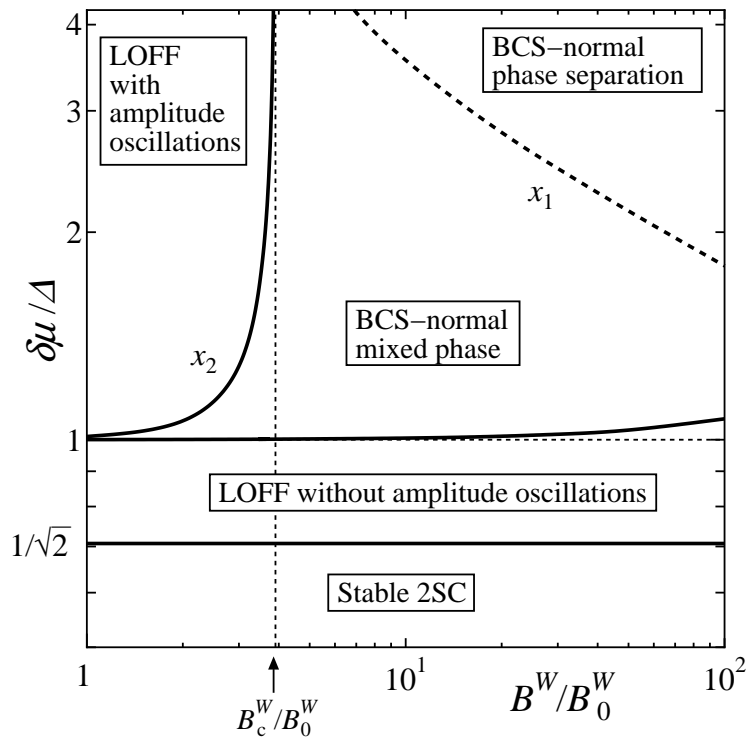}
\end{center}
\caption{
The possible phase structures implied by a stability analysis on
the $B^W$ versus $\delta\mu/\Delta$ plane.
The parameter $B^W$ is normalized by a typical value $B^W_0=50$~MeV~fm$^5$.
The solid line slightly above $\delta\mu/\Delta=1$
represents $v(Q)=0$, while $\delta\mu/\Delta=1$ corresponds to $v_0=0$.
It is thus obvious that the bulk part $v_0$ given by Eq.~(\ref{v02}) 
dominates the behavior of $v(Q)$.  The line labeled with $x_2$ is 
determined by the condition $v_g=0$ from Eq.~(\ref{vg}).  The dashed line 
labeled with $x_1$ indicates the criterion $Q^2=k_{\rm TF}^2$, in the left 
(right) of which the system tends towards a mixed (separated) state.  
Note that with increasing $\delta\mu/\Delta$, $x_2$ and $x_1$ asymptotically 
approach a critical line, $B^W=B^W_c$, given by Eq.\ (\ref{Bcrit}) 
from left and right, respectively.
}
\end{figure}

     The results shown in Fig.~3 are in agreement with our intuitive
expectation:  For larger $B^W$ the BCS-normal interface costs a larger 
energy, meaning that a phase consisting of larger domains is preferable
energetically.  Therefore the region of the BCS-normal phase
separation becomes wider on the phase diagram with increasing $B^W$.  
It should be mentioned that within the present approximation including 
the energy variations of second order in the inhomogeneities,
there is no clear transition between the mixed and separated states;
there can be a first order transition between them, whose
clarification is beyond the scope of this paper.

     Note that we changed $B^W$ as a variable parameter, while
it is equivalent to changing $\Delta$.  This is because the 
conditions determining $x_1$ and $x_2$ would not be altered if 
$\Delta$ is multiplied by an arbitrary factor $\zeta$ and at the
same time $B^W$ is divided by $\zeta^2$.  Thus, a larger $B^W$ with a
fixed $\Delta$ corresponds to a larger $\Delta$ with a fixed $B^W$.
Since $\Delta$ should not exceed $\Delta_{\rm 2SC}$ (see Fig.~1),
however, drastic change in $\Delta$ would not be realistic.

    We conclude this section by noting that the density functional 
(\ref{local}) was originally written by assuming that ${\bf q}$ is small.  
If the true value of ${\bf q}$ is large, therefore, one would need extension 
of the framework by including the higher-order gradient energies.  One
encounters this situation especially when $\delta\mu/\Delta$ is close to 
$x_2$ as can be seen in Fig.~2.

\section{Conclusions}
\label{sec:conc}

    We have systematically examined instabilities of the homogeneous and 
neutral superconducting states for two-flavor quark 
matter with respect to inhomogeneities in the quark and electron densities and
in the phases and amplitude of the order parameter.  The result was summarized
in Fig.\ 3.  We thus clarified the role of the gapless quark modes, the density
and gap gradients, and the electron screening in determining the structure of
spontaneous fluctuations.

    However, open problems still remain.  First, the spatial scale of the 
possible LOFF states remains to be estimated.  This is because this scale is 
determined by the competition between the negative second-order gradient 
term and the unknown fourth-order term \cite{Hatsuda}.  Extension to the case 
of nonzero temperature is significant for possible application to the interiors
of compact stars.  Since the influence of the gapless quark modes on the
thermodynamic potential is smoothed out at nonzero temperature, the 
instabilities as discussed here are likely to be weakened.  Finally, 
extension to the case of three-flavor quark matter is also important for
the description of a more realistic situation.  For this purpose, the 
mean-field analysis of the phase diagram and the chromomagnetic instabilities 
would be a good starting point.

\acknowledgments

    We are grateful to Tetsuo Hatsuda for helpful discussions.  
We thank Ioannis Giannakis, DeFu Hou, Mei Huang, and Hai-cang Ren 
for calling our attention to erroneous expressions in 
Eq.\ (\ref{omegann}) in the earlier version of the manuscript.
This work was supported in part by RIKEN Special Postdoctoral Researchers 
Grant No.\ F86-61016.

\end{document}